\documentclass[11pt]{article}

\usepackage{DGfest,epsfig}
\usepackage{amsmath,amsfonts}
\usepackage{times}
\usepackage{graphicx}
\graphicspath{{./Figures/}}

\newcommand{\etal}{\textit{\mbox{et al.\ }}}          % et al.
\newcommand{\ie}{\textit{\mbox{i.e.\ }}}              % i.e.
\newcommand{\vs}{\textit{\mbox{vs.\ }}}               % vs.

\renewcommand{\Re}{\mbox{Re}\,}                       % Re
\newcommand{\Tr}{\mbox{Tr}}                           % Tr
\newcommand{\bc}{\texttt{bc}}                         % bc
\newcommand{\fc}{\texttt{fc}}                         % fc
\newcommand{\Fig}[1]{Fig.~\ref{#1}}
\newcommand{\Tab}[1]{Table~\ref{#1}}

\newcommand{\Eq}[1]{Eq.~(\ref{#1})}

\newcommand{\beq}{\begin{equation}}
\newcommand{\eeq}{\end{equation}}
\newcommand{\bea}{\begin{eqnarray}}
\newcommand{\eea}{\end{eqnarray}}
\newcommand{\beas}{\begin{eqnarray*}}
\newcommand{\eeas}{\end{eqnarray*}}

\begin{document}
\baselineskip 11.5pt
\title{GAUGE-VARIANT PROPAGATORS AND THE RUNNING COUPLING FROM LATTICE QCD}
%--------------------------------------------------------------------------

\author{E.-M. ILGENFRITZ, M. M{\"U}LLER-PREUSSKER, A. STERNBECK}
\address{Humboldt-Universit{\"a}t zu Berlin, Institut f{\"u}r Physik,
Newtonstr. 15, \\ D-12489 Berlin, Germany}

\author{A. SCHILLER}
\address{Universit{\"a}t Leipzig, Institut f{\"u}r Theoretische Physik, 
Vor dem Hospitaltore 1, \\ D-04103 Leipzig, Germany}

\maketitle
\abstracts{
%---------
On the occasion of the 70's birthday of Prof.~Adriano Di Giacomo
we report on recent numerical computations of the Landau gauge gluon and 
ghost propagators as well as of a non-symmetric MOM-scheme ghost-gluon 
vertex in quenched and full lattice QCD. Special emphasis is paid to the 
Gribov copy problem and to the unquenching effect. The corresponding 
running coupling $~\alpha_s(q^2)~$ is found and shown to decrease for 
$~q^2 \le 0.3 ~\mbox{GeV}^2~$ in the infrared limit. No indication for 
a non-trivial infrared fixed point is seen in agreement with findings 
from truncated systems of Dyson-Schwinger equations treated on a 
four-dimensional torus.
}

\section{Introduction}
%---------------------
\label{sec:intro}
Three of us (E.-M.I., M.M.-P. and A.S.) had the occasion to meet Adriano 
Di Giacomo many times since the late eighties. Over the years we shared 
with him the scientific interest on topological aspects in connection 
with the confinement property of Yang-Mills theories. For us it was 
always helpful and exciting to listen to and to discuss with him and 
to follow his ideas and his opinion. We would like to wish him all the 
best and to be continuously active in science in the future. 

In our contribution to this honorary collection of papers we would like to 
report on a field theoretic investigation closely related to the 
confinement problem which became popular over the last decade. 
It considers the infrared behaviour of the Landau gauge gluon and ghost 
propagators in QCD in connection with the Gribov-Zwanziger horizon 
condition \cite{Zwanziger:1993dh,Gribov:1977wm} and the Kugo-Ojima
criterion \cite{Kugo:1979gm}. The propagators also provide a nonperturbative 
determination of the running QCD coupling $~\alpha_s(q^2)~$ in a MOM scheme. 
It is a great advantage to be able to compute the relevant Green functions 
within the framework of truncated systems of Dyson-Schwinger (DS) equations 
as well as within the lattice approach and to check the consistency of both
approaches. As we shall discuss below, for the infinite volume case 
there is still a mismatch between the infrared limit results, what 
requires further investigations in the near future.

In the infinite 4-volume limit the DS approach has led to an intertwined 
infrared power behaviour of the gluon and ghost dressing functions 
\cite{vonSmekal:1997is,vonSmekal:1998,Alkofer:2000wg,Zwanziger:2001kw,%
Lerche:2002ep} 
\bea \nonumber
Z_{D}(q^2) &\equiv& q^2 D(q^2) \propto (q^2)^{2\lambda}\,, \\
Z_{G}(q^2) &\equiv& q^2 G(q^2) \propto (q^2)^{-\lambda} 
\label{eq:infrared-behaviour}
\eea
with the same $~\lambda\approx 0.59~$, \ie a vanishing gluon propagator 
$D(q^2)$ in the infrared occurs in intimate connection with a diverging 
ghost propagator $G(q^2)$. Under the condition that the ghost-gluon 
vertex renormalization function $Z_1(\mu^2)$ is finite and constant 
(see \cite{Taylor:1971ff} for the perturbative proof and 
\cite{Cucchieri:2004sq} for a first $SU(2)$ lattice study) the corresponding 
MOM scheme running coupling is determined by
\beq
  \alpha_s(q^2) = \frac{g^2}{4\pi}~ Z_{D}(q^2)~ Z^2_{G}(q^2)\,.
\label{eq:runcoupling}
\eeq
This means that \Eq{eq:infrared-behaviour} leads to a non-trivial fixed point 
of $\alpha_s$ in the infrared limit \cite{Lerche:2002ep}.
As we shall show below, our lattice simulations do not agree with this
prediction. 
The strong coupling $\alpha_s(q^2)$ seems to tend to zero in this limit as 
lattice results for other definitions of the running coupling also indicate
(see \cite{Boucaud:1998bq,Boucaud:2002fx} for the three-gluon vertex
and \cite{Skullerud:2002ge} for the quark-gluon vertex). 

Possible solutions of the discrepancy have been discussed from different 
points of view in the recent literature. In \cite{Fischer:2002eq,Fischer:2005} 
the DS approach has been studied on a finite 4-torus with the same 
truncated set of equations as for the infinite volume. 
$~\alpha_s(q^2)~$ was shown to tend to zero for $~q^2 \to 0$ in one-to-one 
correspondence with what one finds on the lattice. This would indicate very 
strong finite-size effects and a slow convergence to the infinite-volume 
limit. However, on the lattice we do not find any indication for such a 
strong finite-size effect, except the convergence to the infinite-volume 
limit would be extremely slow. An alternative resolution of the problem 
has been proposed by Boucaud \etal \cite{Boucaud:2005ce}. These authors 
argued the ghost-gluon vertex in the infrared might contain $q^2$-dependent 
contributions which could modify the DS results for the mentioned 
propagators~~\footnote{We thank A. Lokhov for bringing us the arguments 
in \cite{Boucaud:2005ce} to our attention.}. However, recent detailed DS 
studies of the ghost-gluon vertex did not provide hints for such a 
modification \cite{Schleifenbaum:2004id,Alkofer:2004it}. Thus, at present 
there seems to be no solution of the puzzle. 

For $SU(2)$ extensive lattice investigations can be found in 
\cite{Bloch:2003sk}. Unfortunately, the authors did not reach the 
interesting infrared region, where the mentioned inconsistencies become visible.
The same holds for previous $SU(3)$ lattice computations of the ghost and 
gluon propagators as reported in \cite{Suman:1995zg,Furui:2003jr,Furui:2004cx}.
We have been pursuing an analogous study for the $SU(3)$ case with 
special emphasis on the Gribov copy problem \cite{Sternbeck:2005tk}. 
Moreover, we have investigated the spectral properties of the Faddeev-Popov 
operator \cite{Sternbeck:2005et}. The low-lying eigenmodes of the latter 
are expected to be intimately related to a diverging ghost propagator. 
In addition, in \cite{Sternbeck:2005qj} we have reported on a first
$SU(3)$ lattice computation of the ghost-gluon vertex at 
zero gluon momentum. We confirm, what has been found already for $SU(2)$, 
namely that the data are quite consistent with a constant vertex 
function \cite{Cucchieri:2004sq}. 

For the given report we have extended our investigations restricted 
so far to pure $SU(3)$ Yang-Mills theory to the full QCD case 
with $N_f=2$ clover-improved Wilson fermions. For the latter case we have 
used lattice field configurations produced by the QCDSF collaboration 
and made available via the International Lattice DataGrid 
(ILDG)~\footnote{We thank Gerrit Schierholz and the QCDSF members for
giving us access to their configurations.}. 

\section{Landau gauge fixing, lattice gluon and ghost propagators} 
\label{sec:definitions}
%----------------------------------------------------------------

As a first step we have studied the gluon and ghost propagators in the
quenched approximation. $SU(3)$ gauge field 
configurations \mbox{$U=\{U_{x,\mu}\}$} thermalized with the standard 
Wilson gauge action have been put into the Landau gauge by iteratively 
maximizing the gauge functional 
\beq
  F_{U}[g] = \frac{1}{4V}\sum_{x}\sum_{\mu=1}^{4}\Re\Tr \;{}^{g}U_{x,\mu}\,,
  \qquad 
  {}^{g}U_{x,\mu}=g_x\, U_{x,\mu}\,g^{\dagger}_{x+\hat{\mu}}
  \label{eq:functional}
\eeq
with $g_x \in SU(3)$. It has numerous local maxima (Gribov copies), 
each satisfying the lattice Landau gauge condition 
\beq
  (\partial_{\mu}{}^{g}\!\!A_{\mu})(x) \equiv {}^{g}\!\!A_{\mu}( x+
  \hat\mu/2)-{}^{g}\!\!A_{\mu}( x- \hat\mu/2) = 0
  \label{eq:transcondition}
\eeq 
for the gauge transformed lattice potential 
\beq 
  {}^{g}\!\!A_\mu(x+\hat{\mu}/2) = \frac{1}{2i}\left(^{g} U_{x,\mu} -
                    \ ^{g} U^{\dagger}_{x,\mu}\right)\Big|_{\rm
                    traceless}.
\label{eq:A-definition}
\eeq
To explore to what extent this ambiguity has a significant influence on gauge
dependent observables, we have gauge-fixed each thermalized configuration
a certain number of times (several tens depending on lattice size and coupling) 
mostly using the \emph{over-relaxation} algorithm, starting 
from random gauge copies. For each configuration $U$, we have selected the
first (\fc{}) and the best (\bc{}) gauge copy (that with the largest
functional value) for subsequent measurements. We have checked that 
for the \bc{} copy the number of trials was sufficient for a convergence 
of the ghost propagator within the statistical noise. For details we refer 
to \cite{Sternbeck:2005tk}. 

It turned out that the Gribov ambiguity has no systematic influence on
the infrared behaviour of the gluon propagator. This holds as long as
one restricts to periodic gauge transformations on the four-torus. 
In \cite{Bogolubsky:2005wf} the gauge orbits have been extended to the full 
gauge symmetry including also non-periodic ${\bf Z}(N)$ transformations. 
As a consequence the Gribov problem appears to be enhanced even for the 
gluon propagator. Here we restrict ourselves to the standard case of 
periodic gauge transformations. In marked contrast to the weak Gribov 
copy dependence of the gluon propagator the ghost propagator at lower 
momenta depends on the selection of gauge copies. (For a study of the 
influence of Gribov copies on the ghost propagator in the
$SU(2)$ case see \cite{Bakeev:2003rr}.)

The gluon propagator $D^{ab}_{\mu\nu}(q^2)$ is the Fourier transform
of the gluon two-point function, \ie the colour-diagonal expectation
value
\bea
D^{ab}_{\mu\nu}(q) &=& 
\left\langle \widetilde{A}^a_{\mu}(k)\widetilde{A}^b_{\nu}(-k) \right\rangle \\ 
&=& \delta^{ab} \left( \delta_{\mu\nu} - \frac{q_{\mu}~q_{\nu}}{q^2} \right) 
    D(q^2)\,, \nonumber
\label{eq:D-definition}
\eea
where $\widetilde{A}^a_{\mu}(k)$ denotes the Fourier transform of 
$A^a_\mu(x+\hat{\mu}/2)$ and $q$ is the physical momentum
\beq
  q_{\mu}(k_{\mu}) = \frac{2}{a} \sin\left(\frac{\pi
      k_{\mu}}{L_{\mu}}\right)
\label{eq:p-definition}
\eeq
related to the integer-valued lattice momentum 
$k_{\mu}\in \left(-L_{\mu}/2, L_{\mu}/2\,\right]$ for the linear 
lattice extension $L_{\mu}, \mu=1,\ldots, 4$. According to 
Ref.~\cite{Leinweber:1998uu}, a subset of possible lattice momenta 
$k$ has been chosen for the final analysis of the gluon propagator, 
although the Fast Fourier Transform algorithm provides us with all
lattice momenta. In what follows for the quenched case we use the
lattice spacing $a$ in physical units as determined in \cite{Necco:2001xg}.

The ghost propagator is derived from the Faddeev-Popov (F-P) operator,
the Hessian of the gauge functional given in \Eq{eq:functional}. We
expect that the properties of the F-P operator differ for the
different maxima of the functional (Gribov copies). This has
consequences for the ghost propagator as is shown below.
The F-P operator can be written in terms of the
(gauge-fixed) link variables $U_{x,\mu}$ as
\bea
  M^{ab}_{xy} & = & \sum_{\mu} A^{ab}_{x,\mu}\,\delta_{x,y}
  - B^{ab}_{x,\mu}\,\delta_{x+\hat{\mu},y}
  - C^{ab}_{x,\mu}\,\delta_{x-\hat{\mu},y}\quad
  \label{eq:FPoperator}
\eea
with
\beas
  A^{ab}_{x,\mu} &=& \phantom{2\cdot\ } \Re\Tr\left[
    \{T^a,T^b\}(U_{x,\mu}+U_{x-\hat{\mu},\mu}) \right],\\
  B^{ab}_{x,\mu} &=& 2\cdot\Re\Tr\left[ T^bT^a\, U_{x,\mu}\right],\\
  C^{ab}_{x,\mu} &=& 2\cdot\Re\Tr\left[ T^aT^b\, U_{x-\hat{\mu},\mu}\right]
\eeas
and $~T^a,~a=1,\ldots,8~$ being the (hermitian) generators of the
$~\mathfrak{su}(3)~$ Lie algebra satisfying $~\Tr~[T^aT^b]~=~\delta^{ab}/2$.
The ghost propagator is then determined by inverting the F-P operator $~M$ 
\bea
  G^{ab}(q) &=& \frac{1}{V} \sum_{x,y} \left\langle {\rm e}^{-2\pi i\,k
      \cdot (x-y)} [M^{-1}]^{ab}_{xy} \right\rangle_U\,, \\
            &=& \delta^{ab} G(q^2) \,. \nonumber
\label{eq:Def-ghost}
\eea
Following Ref.~\cite{Suman:1995zg,Cucchieri:1997dx} we have used the 
conjugate gradient (CG) algorithm to invert $M$ on a plane wave $\vec{
\psi}_c$ with colour and position components \mbox{$\psi^a_c(x) = \delta^{ac}
\exp (2\pi i\,k\!\cdot\! x)$}. In fact, we applied the pre-conditioned
CG algorithm (PCG) to solve $M^{ab}_{xy}\phi^{b}(y)=\psi^a_c(x)$. As 
the pre-conditioning matrix we used the inverse Laplacian operator 
$\Delta^{-1}$ with diagonal colour substructure. This has significantly 
reduced the required amount of computing time (for details see 
\cite{Sternbeck:2005tk}). 
After solving $M\vec{\phi}=\vec{\psi}_c$ the resulting
vector $\vec{\phi}$ is projected back on $\vec{\psi}_c$
such that the average $G^{cc}(q)$ over the colour index $c$ (divided by $V$) 
can be taken explicitly. Since the F-P operator $M$ is singular if acting on
constant modes, only $k \ne (0,0,0,0)$ is permitted. Due to high
computational requirements to invert the F-P operator for each $k$,
separately, the estimator on a single, gauge-fixed configuration is
evaluated only for a preselected set of momenta $k$. 

\section{Quenched QCD results}
%-----------------------------
\label{sec:quenchedresults}

%--------------------------------------------------------------------- 
\begin{figure}[ht]
\includegraphics[width=6.5cm,height=6.0cm]{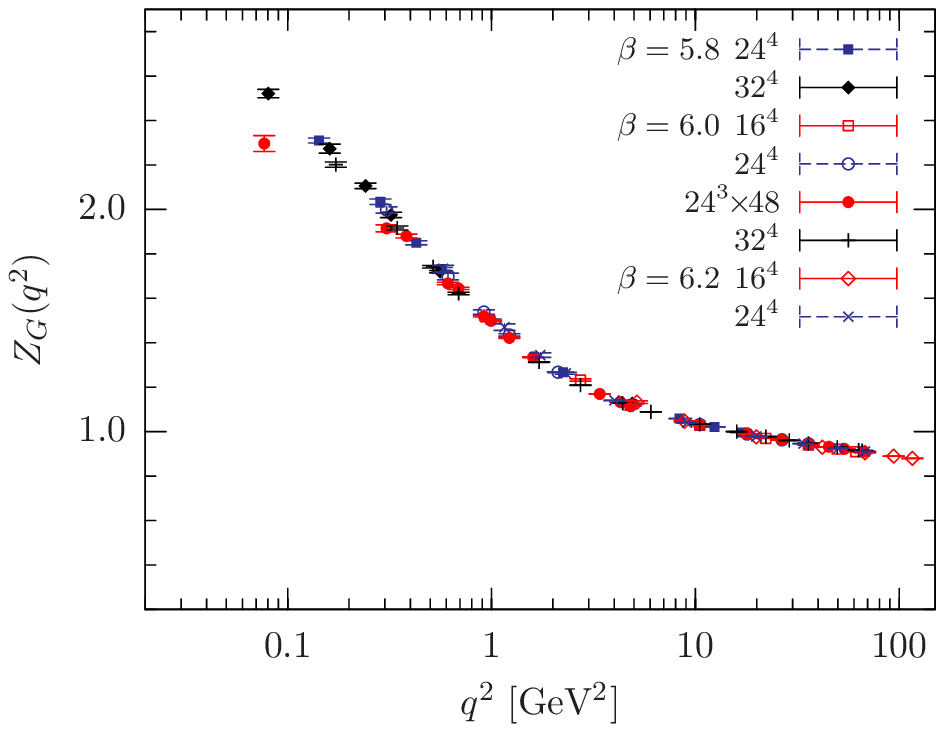} 
\quad
\includegraphics[width=6.5cm,height=6.0cm]{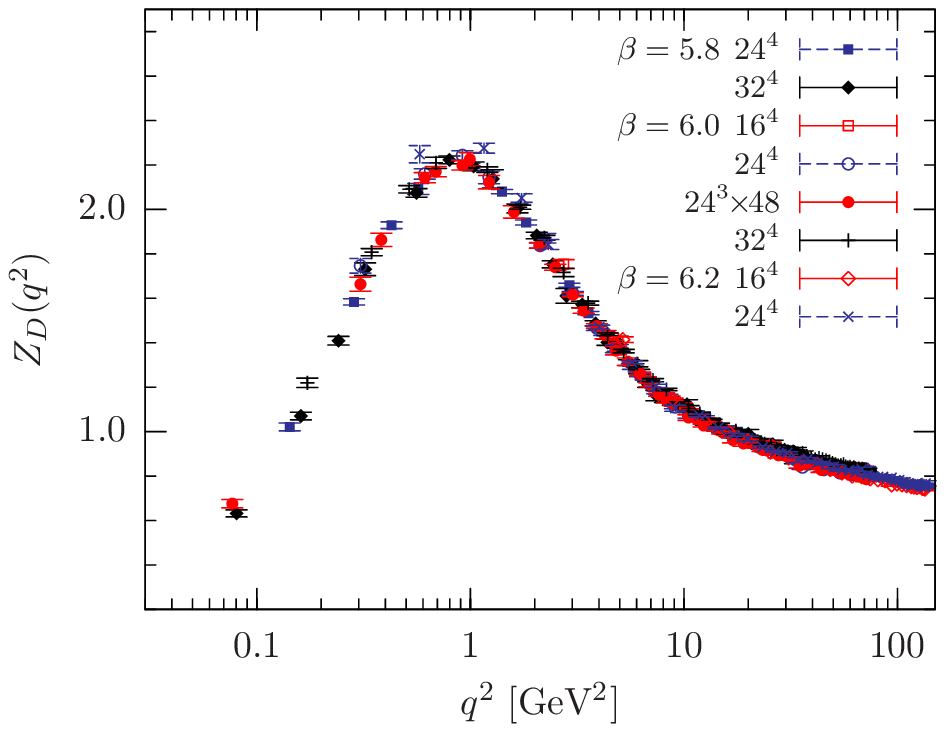} 
\caption{
The dressing functions for the ghost $~Z_G \equiv q^2 G(q^2)~$ (l.h.s.) 
and gluon propagator $~Z_D \equiv q^2 D(q^2)~$ (r.h.s.) vs. $q^2$
for quenched QCD, both measured on \fc{} gauge copies. 
}
\label{fig:ghost_gluon}
\end{figure}
%---------------------------------------------------------------------

%--------------------------------------------------------------------- 
\begin{figure}[ht]
\includegraphics[width=6.5cm,height=4.5cm]{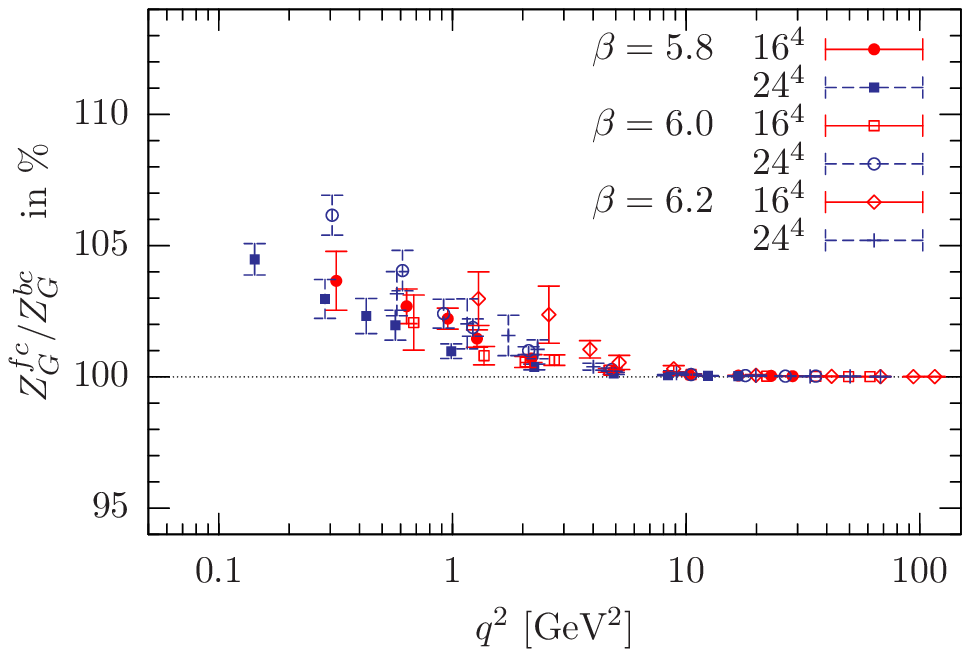} 
\quad
\includegraphics[width=6.5cm,height=4.5cm]{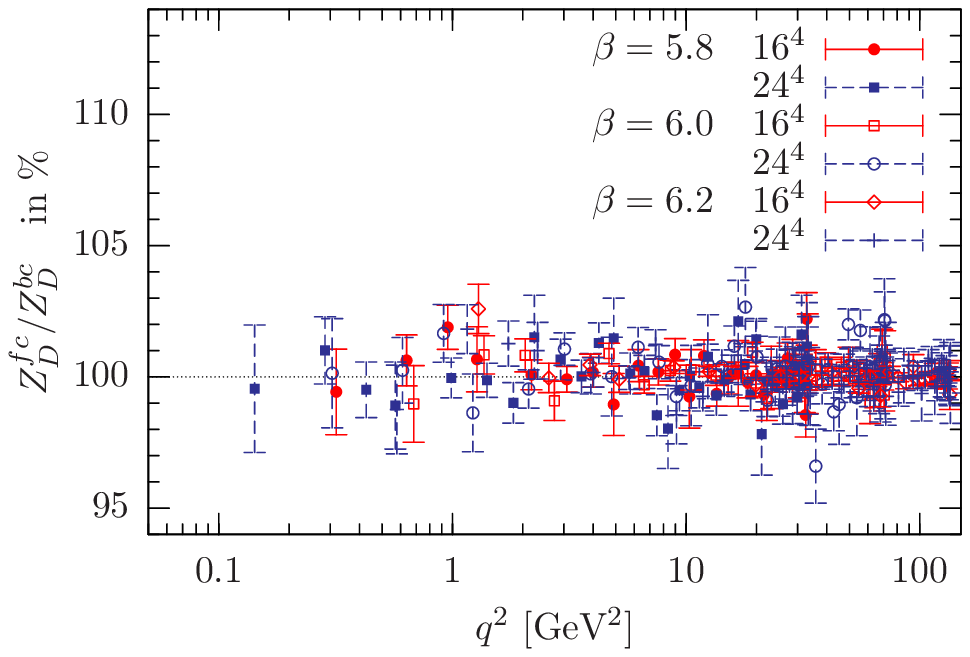}
\caption{
The ratios $Z^{\rm fc} / Z^{\rm bc}$ for the dressing functions 
$Z_G$ (l.h.s.) and $Z_D$ (r.h.s.) determined on first (\fc{}) 
and best (\bc{}) gauge copies \vs momentum $q^2$.
}
\label{fig:fcbc_ratios}
\end{figure}
%---------------------------------------------------------------------

In \Fig{fig:ghost_gluon} we show the ghost and the 
gluon dressing functions $Z_{G,D}(q^2)$ versus $q^2$ for \fc{} copies 
in the present state of our quenched QCD simulations.
Both propagators have been renormalized separately for each $\beta$ 
with the normalization condition $Z_{G,D}=1$ at $q  = 4~\mbox{GeV}$.

In \Fig{fig:fcbc_ratios} we illustrate the effect of the 
Gribov copies for periodic gauge transformations.  
We have plotted some \fc{} - to - \bc{} ratios of the ghost and gluon dressing 
functions. Obviously there is no influence visible for the gluon propagator
within the statistical noise. On the contrary, for the ghost propagator the 
Gribov problem can cause $O(5 \%)$ deviations in the low-momentum region 
($q<1~\mbox{GeV}$). For better gauge copies the ghost dressing function 
becomes less singular in the infrared. A closer inspection of the data 
for the ghost propagator indicates that the influence of Gribov copies 
becomes weaker for {\bf increasing} lattice size. This is in agreement 
with a recent claim by Zwanziger according to which in the infinite volume 
limit averaging over gauge copies in the Gribov region should lead to 
the same result as averaging over copies restricted to the fundamental modular 
region \cite{Zwanziger:2003cf}. In \cite{Bogolubsky:2005wf} similar 
indications have been found for $SU(2)$ taking non-periodic
$Z(2)$ transformations into account.
 
Note in \Fig{fig:ghost_gluon} the deviation of the lowest-momentum 
$Z_G$ data point obtained on an asymmetric lattice $24^3 \times 48$. 
This deviation deserves further study and sheds some critical light 
on investigations on strongly asymmetric lattices 
\cite{Silva:2004bv,Oliveira:2004gy}. Still it remains difficult to 
extract a possible power behaviour at low $q^2$ as in 
\Eq{eq:infrared-behaviour}.
 
\begin{figure}[ht] 
\includegraphics[width=6.5cm,height=6.0cm]{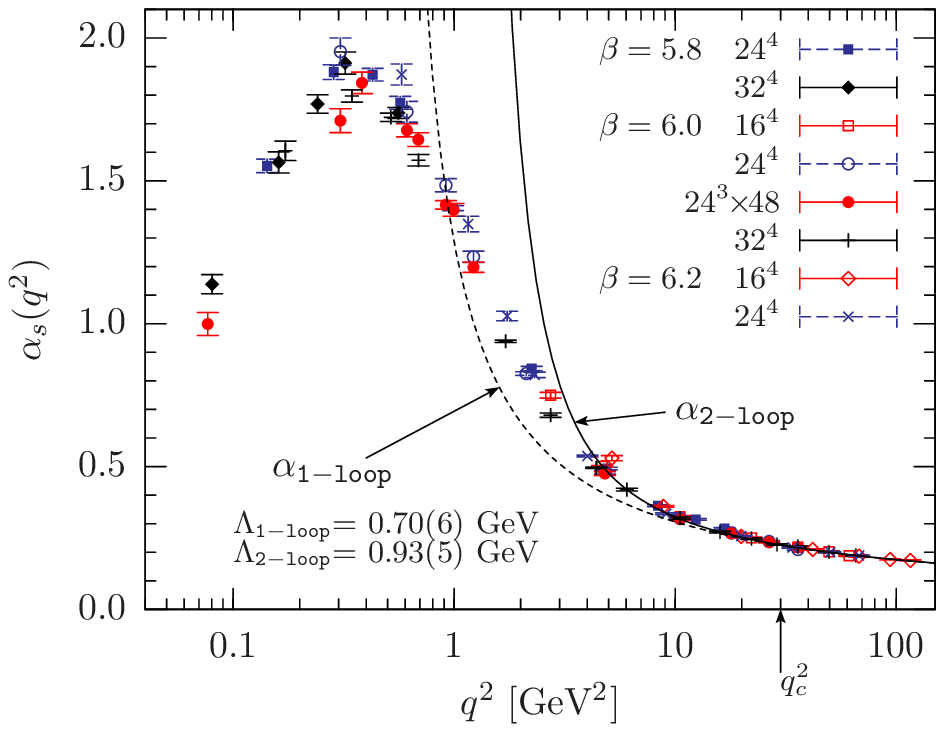} \quad
\includegraphics[width=6.5cm,height=6.0cm]{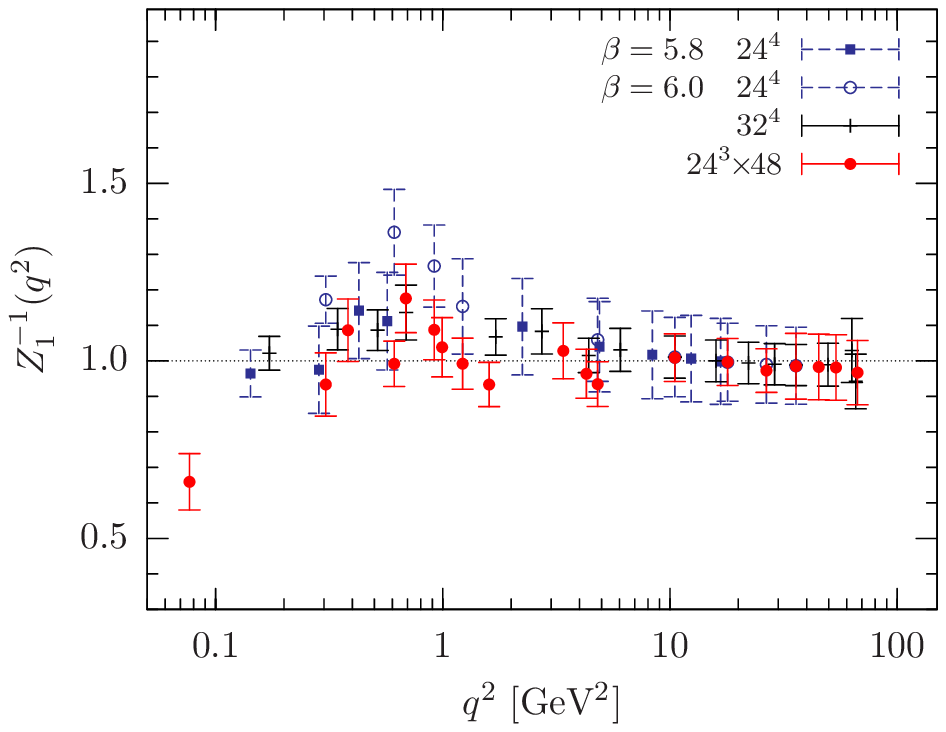}
\caption{
L.h.s.: The momentum dependence of the running coupling $\alpha_{s}(q^2)$ 
for quenched QCD measured on \fc{} gauge copies.
R.h.s.: The inverse ghost-gluon vertex renormalization function 
$Z_1^{-1}(q^2)$ measured on \fc{} gauge copies and normalized to unity 
at $q=4~\mbox{GeV}$.
} 
\label{fig:alpha}
\end{figure}

In \Fig{fig:alpha} we present the combined result for the running coupling 
according to \Eq{eq:runcoupling} for \fc{} copies in the quenched QCD 
case. Fits to the 1-loop and 2-loop running coupling are also shown.
%---------------------------------------------------------------------
%---------------------------------------------------------------------
Our data confirm the running coupling monotonously to decrease 
with decreasing momentum in the range $q^2 < 0.3~\mbox{GeV}^2$. The nice
coincidence of the data for various lattice sizes (except the points 
refering to the asymmetric lattice) makes it hard to see how finite-volume 
effects can be blamed for the disagreement with the DS continuum result 
\Eq{eq:infrared-behaviour}.    
  
One can ask, whether the ghost-gluon vertex renormalization function 
$Z_1(q^2)$ is really constant at lower momenta. A recent investigation 
of this function defined at vanishing gluon momentum for the $SU(2)$ case 
\cite{Cucchieri:2004sq} supports that \mbox{$Z_1(q^2)\approx 1$} at least 
for momenta larger than 1~GeV. We have performed an analogous study for 
$Z_1(q^2)$ in the case of $SU(3)$ gluodynamics. Our results are presented 
on the right hand side of \Fig{fig:alpha}. There is a slight variation
visible in the interval $0.3 ~\mbox{GeV}^2 \le q^2 \le 5 ~\mbox{GeV}^2$.
But this weak deviation from being constant will not have a dramatic 
influence on the running coupling. There is a data point at the lowest 
available momentum derived from the asymmetric lattice which deviates in the
opposite direction. Current simulations on larger lattices will enable us 
to draw conclusions in the near future.

\section{Full QCD results}
%-------------------------
\label{sec:fullresults}

Now let us turn to the full QCD case. We have investigated ILDG gauge field 
configurations generated in simulations by the QCDSF collaboration
\cite{Gockeler:2004rp,Gockeler:2005rv}. $N_f=2$ flavours of dynamical 
clover-improved Wilson fermions together with the Wilson plaquette gauge 
action have been used. The (asymmetric) lattice size is $24^3\times 48$ 
possibly demanding some caution in the very infrared as discussed already
in the quenched case. The values of the bare coupling $\beta$ and of the 
quark mass $ma$ 
\beq
\label{eq:ma}
  ma = \frac{1}{2}\left(\frac{1}{\kappa}-\frac{1}{\kappa_c}\right)
\eeq
taken into account are collected in \Tab{tab:stat}.
The values for $\kappa_c$ can be found in \cite{Gockeler:2005rv}. 
The last column of the Table gives the number of configurations investigated.
%The corresponding Markov-Chain-URI of the QCDSF/ILDG configurations are: \\
%{\tt www.lqcd.org/ildg/qcdsf/b5p25kp13575-24x48}, \\
%{\tt www.lqcd.org/ildg/qcdsf/b5p29kp13550-24x48} and \\ 
%{\tt www.lqcd.org/ildg/qcdsf/b5p29kp13590-12x32}.
The lattice spacing has been translated into physical units 
via the Sommer scale $r_0/a$ as determined in \cite{Gockeler:2005rv} 
with $r_0=0.5$~fm. For fixing to Landau gauge we have employed the 
over-relaxation or the Fourier-accelerated gauge-fixing method. Again
we concentrate on \fc{} copy results postponing the question of the
influence of Gribov copies in the full QCD case to a future publication. 

%---------------------------------------------------------------------------
\begin{table}[t]
\caption{
The parameter values $\beta$, $\kappa$ etc. used for our investigation. 
}
\begin{center}
\begin{tabular}{|c|c|c|c|c|c|}
\hline
$\beta$ & $\kappa$ & $\kappa_c$ & $ma$ & $a [\mbox{GeV}^{-1}]$ & \#~\mbox{conf} 
\\
\hline
  5.29 & 0.13550 & 0.136410\,(09) & 0.0246 & 2.197 & 60 \\
  5.29 & 0.13590 & 0.136410\,(09) & 0.0138 & 2.324 & 55 \\
  5.25 & 0.13575 & 0.136250\,(07) & 0.0135 & 2.183 & 60 \\
\hline
\end{tabular}
\end{center}
\label{tab:stat}
\end{table}
%---------------------------------------------------------------------------

\begin{figure}[ht]
\includegraphics[width=6.5cm,height=6.5cm]{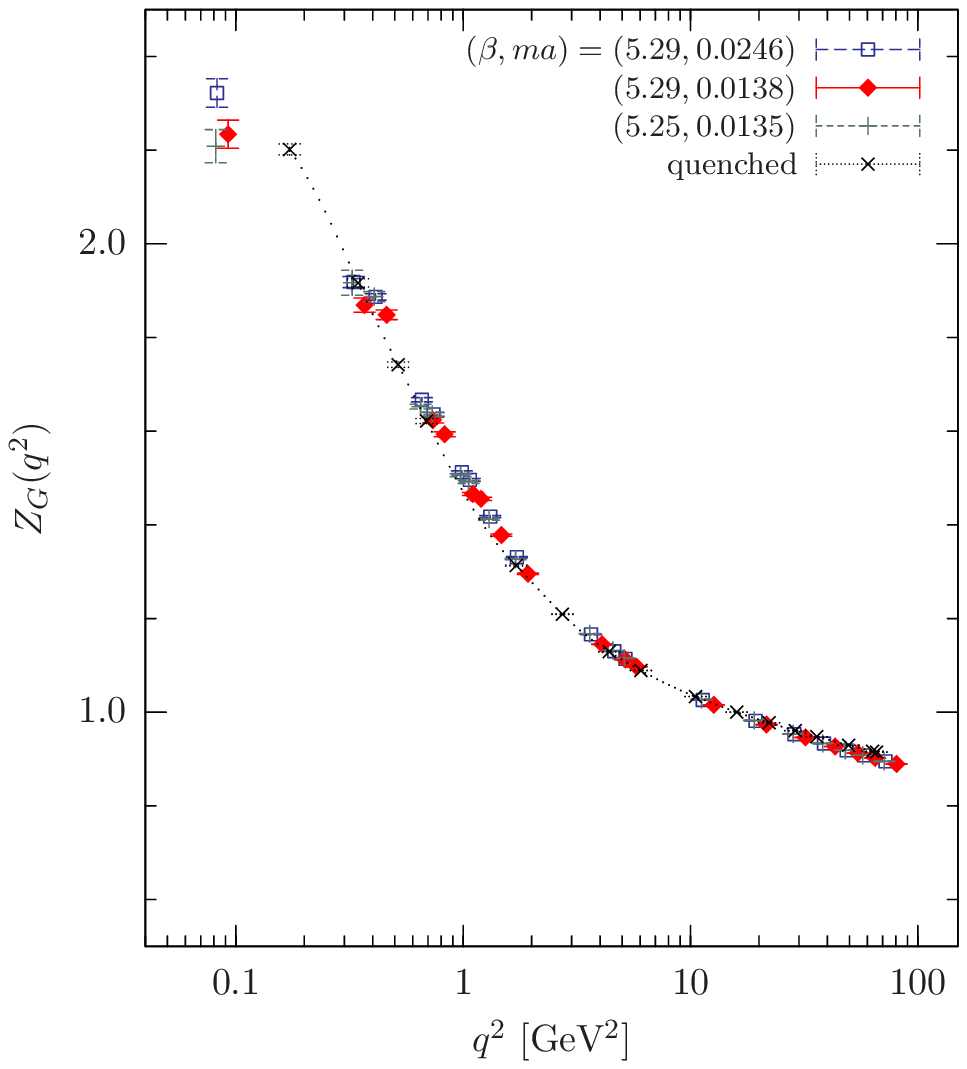} $\quad$
\includegraphics[width=6.5cm,height=6.5cm]{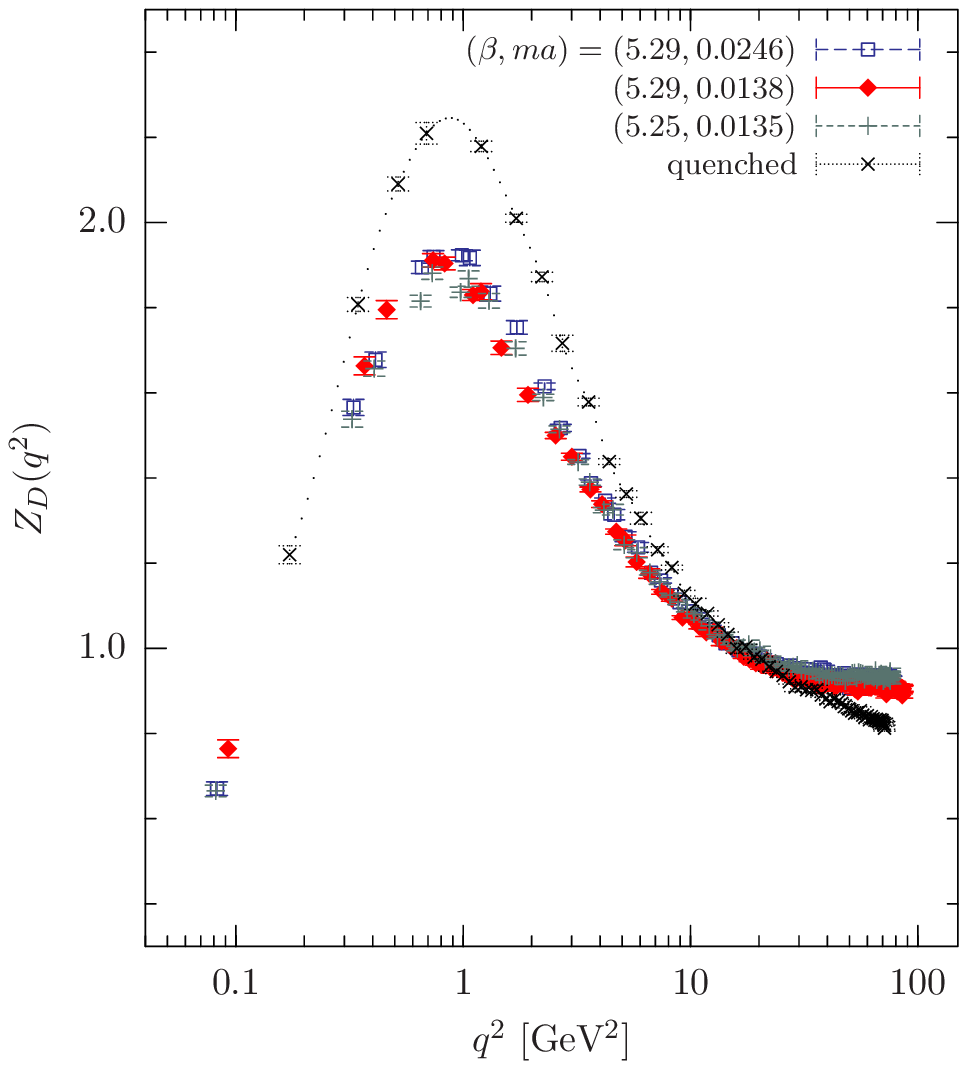}
\caption{
The dressing functions of the ghost propagator $Z_{G}$ (l.h.s.) 
and the gluon propagator $Z_{D}$ (r.h.s.) versus $q^2$ 
for full QCD with $N_f=2$ flavours of clover-improved Wilson 
fermions (lattice size $24^3 \times 48$). 
Quenched QCD data for $\beta=6.0$ and $32^4$ are
shown for comparison. The dressing functions are renormalized to
$Z=1$ at $q=4~\mbox{GeV}$. All the data refer to \fc{} gauge copies.
} 
\label{fig:gh_and_gl_as_func_of_q}
\end{figure}

In \Fig{fig:gh_and_gl_as_func_of_q} we present our full QCD results 
for the ghost and gluon dressing functions and, for comparison, selected data 
points of the quenched case (\i.e. for infinite quark mass)
for $\beta=6.0$ and a $32^4$ lattice. 
All dressing functions have been renormalized such 
that $Z_{G,D}=1$ at $q = 4~\mbox{GeV}$.

%---------------------------------------------------------------------------
%---------------------------------------------------------------------------

The unquenching effect becomes clearly visible for the gluon propagator,
whereas the ghost propagator does not show any quark mass dependence. 
The non-perturbative peak of the gluon propagator at $q \simeq 1~\mbox{GeV}$
becomes softer as the quark mass is decreasing. This has been observed 
also in recent lattice computations of the gluon propagator using 
dynamical AsqTad improved staggered quarks \cite{Bowman:2004jm} as well as
from unqenching studies for the ghost and gluon propagators within the 
DS equation approach \cite{Fischer:2005wx,Fischer:2005en,Fischer:2003rp}.
We refer also to recent lattice studies with dynamical Kogut-Susskind
and Wilson fermions \cite{Furui:2005bu,Furui:2005he}.

\begin{figure}[ht]
\includegraphics[width=6.5cm,height=6.5cm]{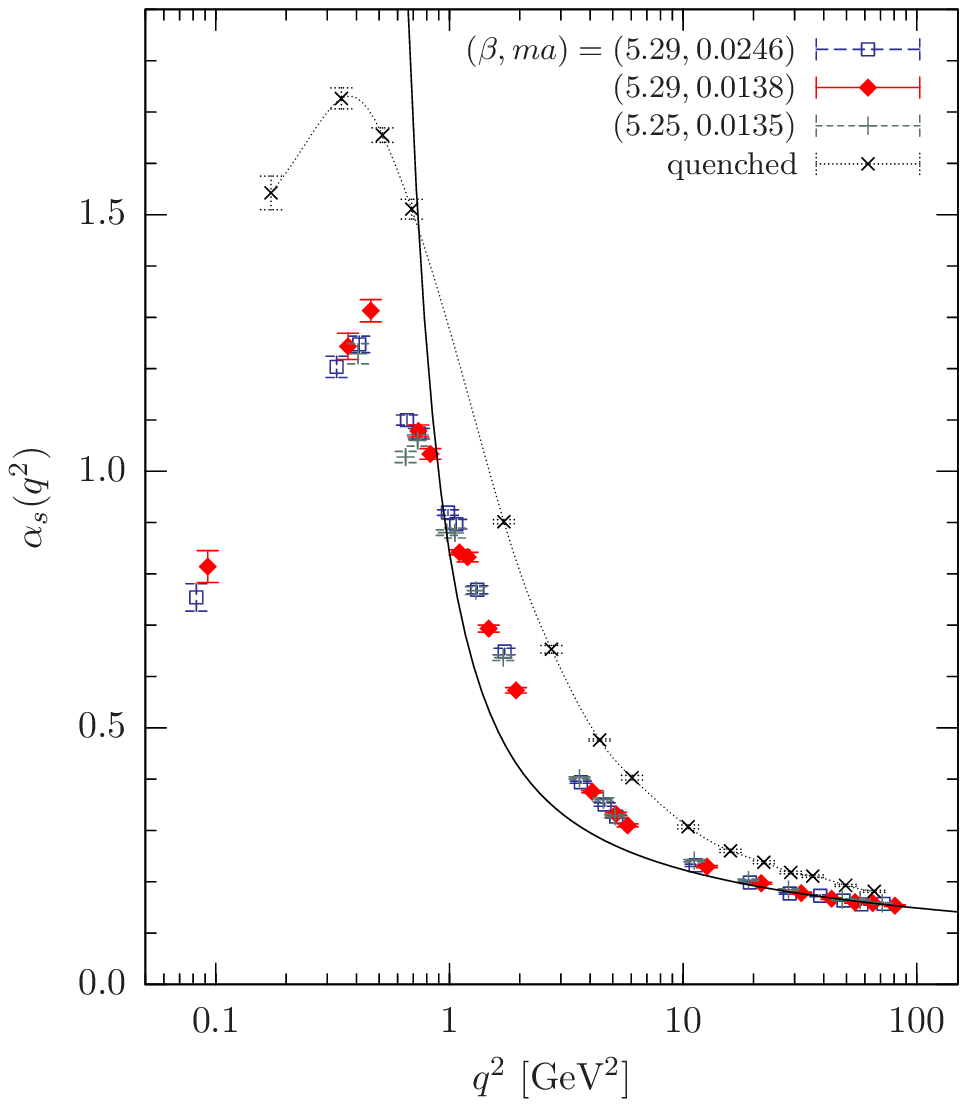}  $\quad$
\includegraphics[width=6.5cm,height=6.5cm]{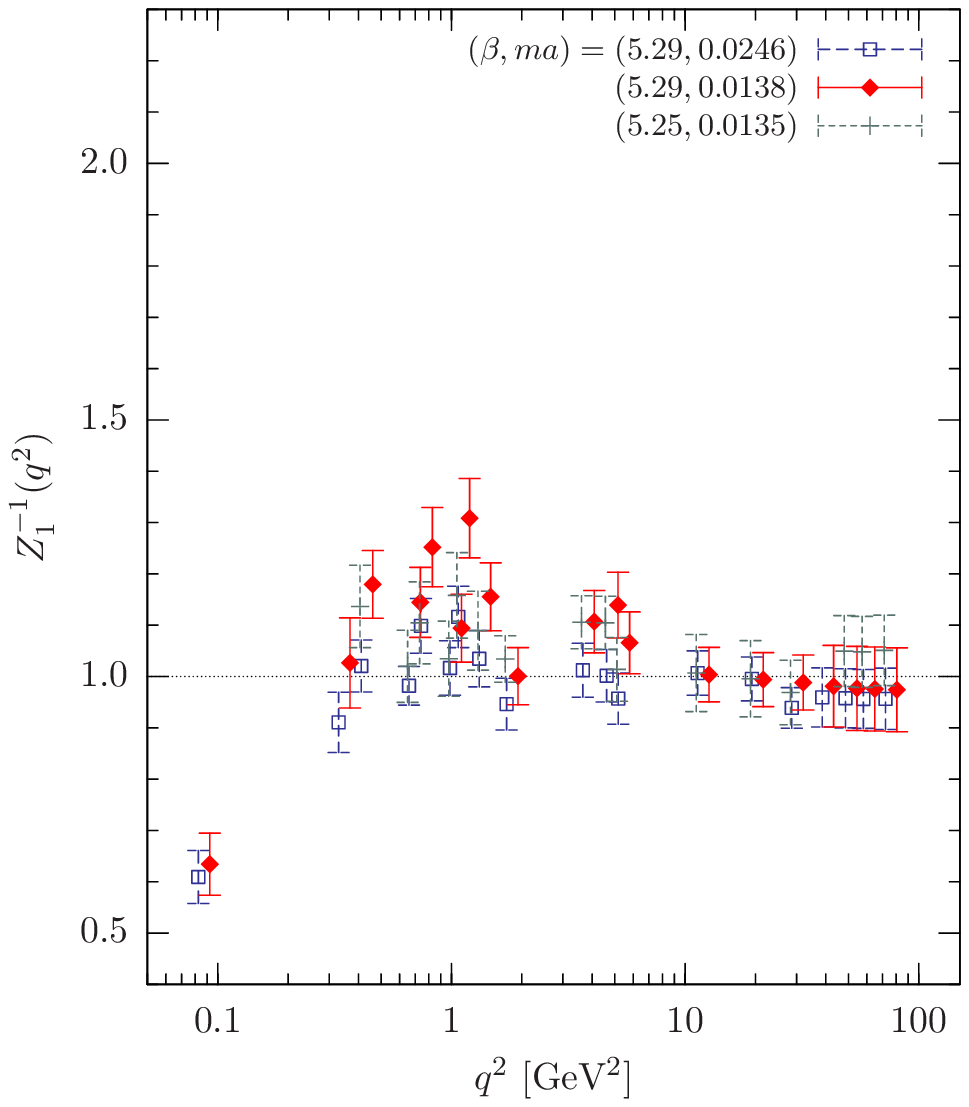}
\caption{
The running coupling (l.h.s.) and the inverse renormalization
function $Z_1^{-1}$ of the ghost-gluon vertex (r.h.s.) are shown as
functions of the physical momentum for full QCD. For comparison, the
quenched QCD running coupling data for $\beta=6.0, ~32^4$ are shown 
with a spline fit line to guide the eye. The 2-loop fit to 
$\alpha_s(q^2)$ for $N_f=2$ is drawn with the solid line. 
The vertex renormalization function is normalized to $Z_1=1$ 
at $q=4~\mbox{GeV}$.
}
\label{fig:alpha_and_z1_as_func_of_q}
\end{figure}

In \Fig{fig:alpha_and_z1_as_func_of_q} we show the corresponding
MOM scheme running coupling $\alpha_s(q^2)$ and the inverse of the
renormalization function $Z_1(q^2)$ of the ghost-gluon vertex 
as functions of the physical momentum. Again we compare results for 
full QCD with those for the quenched case at $\beta=6.0$ on a 
$32^4$ lattice. The normalization factors of the running couplings were
determined by fitting the data to the (massless) 2-loop coupling  
formulae for $\alpha_s(q^2)$ with $N_f=2$ and $N_f=0$, respectively, 
both in the range $q^2>30 ~\mbox{GeV}^2$. The vertex renormalization 
function was normalized to $Z_1=1$ at $q=4~\mbox{GeV}$ for each data set 
($\beta, ma$) separately.

For the running coupling we observe a quite clear unquenching effect 
extending from the perturbative range down to the infrared region. 
Again we see the coupling to turn down in the infrared limit. However, 
the data points at $q^2 \simeq 0.1 \mbox{GeV}^2$ have to be checked 
carefully on symmetric and larger lattices. The same holds for $Z_1$.     

%---------------------------------------------------------------------------
%---------------------------------------------------------------------------

\section{Conclusions}
%--------------------
We have studied the low-momentum region of QCD within the Landau gauge
using Monte Carlo simulations with the Wilson plaquette action 
and various lattice sizes from $16^4$ to $32^4$ for the quenched case
in the range of bare couplings $\beta=5.8, \ldots, 6.2$. In this way 
we have reached momenta down to $q^2 \simeq 0.1 \mbox{GeV}^2$. 
Moreover, for studying unquenching effects we have carried out an
analogous investigation in full QCD for the case of two improved 
Wilson flavours for a couple of quark mass values
but only one lattice size $24^3 \times 48$. Our data presented here
refer to first gauge copies as obtained from over-relaxation or 
Fourier-accelerated gauge fixing. In the quenched QCD case we have seen 
clear Gribov copy effects in the infrared region for the ghost propagator.
The ghost propagator becomes less singular within a $O(5\%)$ deviation, 
when better gauge copies are taken. We have found indications that the 
Gribov effect also weakens as the 
volume is increasing. Towards $q \to 0$ in the infrared momentum region, 
the gluon dressing functions in quenched as well as full QCD were shown to 
decrease, while the ghost dressing functions turned out to rise. 
However, the connected power laws predicted by the infinite-volume 
DS approach could not be confirmed on the basis of our data. Correspondingly, 
the behaviour of the running coupling $\alpha_s(q^2)$ in a suitable 
momentum subtraction scheme (based on the ghost-gluon vertex) does 
not approach the expected finite limit.  Instead, this coupling has been 
found to decrease for lower momenta after passing a turnover at 
$q^2 \approx 0.4$~GeV$^2$ for quenched ($N_f=0$) as well as for full QCD
($N_f=2$). 
Unquenching effects have been clearly recovered for the gluon propagator,
whereas the ghost propagator was almost unchanged. This is in one-to-one
correspondence with what has been recently found in the Dyson-Schwinger 
equation approach. However, the puzzle of the existence of a non-trivial
infrared fixed point in the infinite volume limit remains unsolved. 

\section*{Acknowledgements}
%--------------------------
All simulations have been done on the IBM pSeries 690 at HLRN.  We
are indebted to the QCDSF collaboration for providing us with their 
gauge field configurations via the International Lattice DataGrid (ILDG) 
({\tt http://www.lqcd.org/ildg/}). We used \emph{ltools} for getting those
configurations \cite{Stuben:2005uf}. We thank
Dirk Pleiter and Stefan Wollny for their help as well as H.~St\"uben for
contributing parts of the program code. We also acknowledge discussions
with G. Burgio, C. Fischer and V.K. Mitrjushkin. This work has been 
supported by the DFG under contract FOR~465 (Forschergruppe {\it Lattice 
Hadron Phenomenology}).  A.~Sternbeck acknowledges support of the 
DFG-funded graduate school GK~271.

\section*{References}
%--------------------
%\bibliographystyle{unsrthep}
%\bibliography{references}

\end{document}